\documentstyle[A4,11pt,epsfig,twoside]{article}
\def\ie{{i.e.}}

\def\ra{\rightarrow}

\def\be{\begin{equation}}
\def\ee{\end{equation}}
\def\bea{\begin{eqnarray}}
\def\eea{\end{eqnarray}}
\def\tbtb{\rm t\anti b \, \anti t b}
\def\bbbb{\rm b\anti b b\anti b}
\def\wt{\widetilde}
\def\epem{\rm e^+e^-}
\def\bb{\rm b\anti{b}}

\def\bbA{\rm \bb \ha}
\def\tanb{\tan\beta}

\def\hl{\rm h}
\def\hh{\rm H}
\def\ha{\rm A}
\def\hp{\rm H^+}
\def\hm{\rm H^-}
\def\hpm{\rm H^\pm}
\def\mhpm{m_{\rm \hpm}}
\def\anti{\overline}
\def\mhh{m_{\rm \hh}}
\def\mha{m_{\rm \ha}}

\def\gev{~{\rm GeV}}

\def\fbi{~{\rm fb}^{-1}}

\def\call{{\cal L}}
\def\rts{\sqrt s}

\def\br{{\rm BR}}
\def\gamhatot{\Gamma_{\rm tot}^{\rm \ha}}
\def\gamhhtot{\Gamma_{\rm tot}^{\rm \hh}}
\def\gamhpmtot{\Gamma_{\rm tot}^{\rm \hpm}}
\def\gamres{\Gamma_{\rm res}}
\def\mt{m_{\rm t}}
\def\mb{m_{\rm b}}
\def\lsim{\mathrel{\raise.3ex\hbox{$<$\kern-.75em\lower1ex\hbox{$\sim$}}}}
\def\gsim{\mathrel{\raise.3ex\hbox{$>$\kern-.75em\lower1ex\hbox{$\sim$}}}}
\def\ifmath#1{\relax\ifmmode #1\else $#1$\fi}
\def\half{\ifmath{{\textstyle{1 \over 2}}}}

\def\beq{\begin{equation}}
\def\eeq{\end{equation}}
\def\bit{\begin{itemize}}
\def\eit{\end{itemize}}

\begin{document}
\begin{titlepage}
\def\thefootnote{\fnsymbol{footnote}}       

\begin{center}
\mbox{ } 

\end{center}
\vskip -3.0cm
\begin{flushright}
\Large
\vspace*{-4cm}
\mbox{\hspace{11.85cm} hep-ph/0211123} \\
\mbox{\hspace{12.0cm} November 2002}
\end{flushright}
\begin{center}
\vskip 1.0cm
{\boldmath \Huge\bf
\mbox{Overview of $\tan\beta$ Determination}
\smallskip
\mbox{at a Linear $\rm \epem$ Collider}
}
\vskip 1cm
{\LARGE\bf J. Gunion$^1$, T. Han$^2$, J. Jiang$^3$,
           A. Sopczak$^4$}\\
\smallskip
\smallskip
\smallskip
\Large
$^1$Davis Institute for HEP, U. of California, Davis, CA \\
$^2$Dept. of Physics, U. of Wisconsin, Madison, WI\\
$^3$Argonne National Laboratory, Chicago, IL \\
$^4$Lancaster University, UK

\vskip 1.5cm
\centerline{\Large \bf Abstract}
\end{center}

\vskip 1.cm
\hspace*{-0.5cm}
\begin{picture}(0.001,0.001)(0,0)
\put(,0){
\begin{minipage}{14cm}
\Large
\renewcommand{\baselinestretch} {1.2}
The ratio of the vacuum expectation value of the two Higgs doublets,
$\tanb$, is an important parameter of the general 2-Higgs-Doublet
Model (2HDM) and the Minimal Supersymmetric extension of the Standard
Model (MSSM). The expected uncertainty on the determination of 
$\tanb$ at a Linear Collider (LC) 
of at least 500~GeV center-of-mass energy and high luminosity is
reviewed based on studies of neutral and charged Higgs boson production.
\renewcommand{\baselinestretch} {1.}

\normalsize
\vspace{1cm}
\begin{center}
{\sl \large
Presented at the SUSY02 conference, DESY, June 2002; \\
and the LCWS02, Korea, Aug. 2002
\vspace{-3cm}
}
\end{center}
\end{minipage}
}
\end{picture}
\vfill

\end{titlepage}


\newpage
\thispagestyle{empty}
\mbox{ }
\newpage
\setcounter{page}{1}

\title{OVERVIEW OF $\tan\beta$ DETERMINATION \\ 
       AT A LINEAR $\rm \epem$ COLLIDER}

\author{J. GUNION$^1$, T. HAN$^2$, J. JIANG$^3$, A. SOPCZAK$^4$\footnote{speaker}\\ \\
$^1$Davis Institute for HEP, U. of California, Davis, CA \\
$^2$Dept. of Physics, U. of Wisconsin, Madison, WI\\
$^3$Argonne National Laboratory, Chicago, IL \\
$^4$Lancaster University, UK}

\date{}
\maketitle
\vspace*{-1.0cm}
\begin{abstract}
\vspace*{-0.3cm}
The ratio of the vacuum expectation value of the two Higgs doublets,
$\tanb$, is an important parameter of the general 2-Higgs-Doublet
Model (2HDM) and the Minimal Supersymmetric extension of the Standard
Model (MSSM). The expected uncertainty on the determination of
$\tanb$ at a Linear Collider (LC)
of at least 500~GeV center-of-mass energy and high luminosity is
reviewed based on studies of neutral and charged Higgs boson production.
\end{abstract}

\vspace*{-0.3cm}
\section*{Introduction}
\vspace*{-.6cm}
Various methods to determine $\tanb$ at a LC exist and they
have in common that a physical observable depends on $\tanb$:
\begin{itemize}
\item The pseudoscalar Higgs boson, A, could be produced
      via radiation off a pair of b-quarks:
      $\rm e^+e^-\rightarrow b\bar b\rightarrow b\bar b A 
        \rightarrow b\bar b b\bar b.$
      The $\rm b\bar b A$ coupling is proportional to $\tan\beta$ and thus the 
      expected production rate is proportional to $\tan^2\beta$.
\item The $\rm b\bar b b\bar b$ rate 
      from the pair-production of the heavier scalar, H, in association 
      with the pseudoscalar Higgs boson
      $\rm e^+e^- \rightarrow HA \rightarrow b\bar b b\bar b$
      can be exploited.
      While the HA production rate is almost independent of $\tan\beta$ 
      the sensitivity occurs via the variation of the 
      decay branching ratios with $\tan\beta$. 
\item The value of $\tan\beta$ can also be determined from the 
      H and A decay widths, which can be obtained from the previously
      described reaction.
\item The $\tbtb$ rate from charged Higgs boson production
      can contribute to the 
      determination of $\tan\beta$ from the reaction 
      $\rm \epem\to H^+H^-\to \tbtb$ because of the charged Higgs boson
      branching ratio variation with $\tanb$.
\item In addition, the charged Higgs boson total decay width depends 
      on $\tanb$.  
\end{itemize}

\vspace*{-.6cm}
\section*{The \boldmath $\rm \bbA\to \bb\bb$ bremsstrahlung process}
\vspace*{-6mm}

\begin{figure}[h!]
\begin{minipage}{0.48\textwidth}
\begin{center}
\vspace*{-1.4cm}
\mbox{\epsfig{file=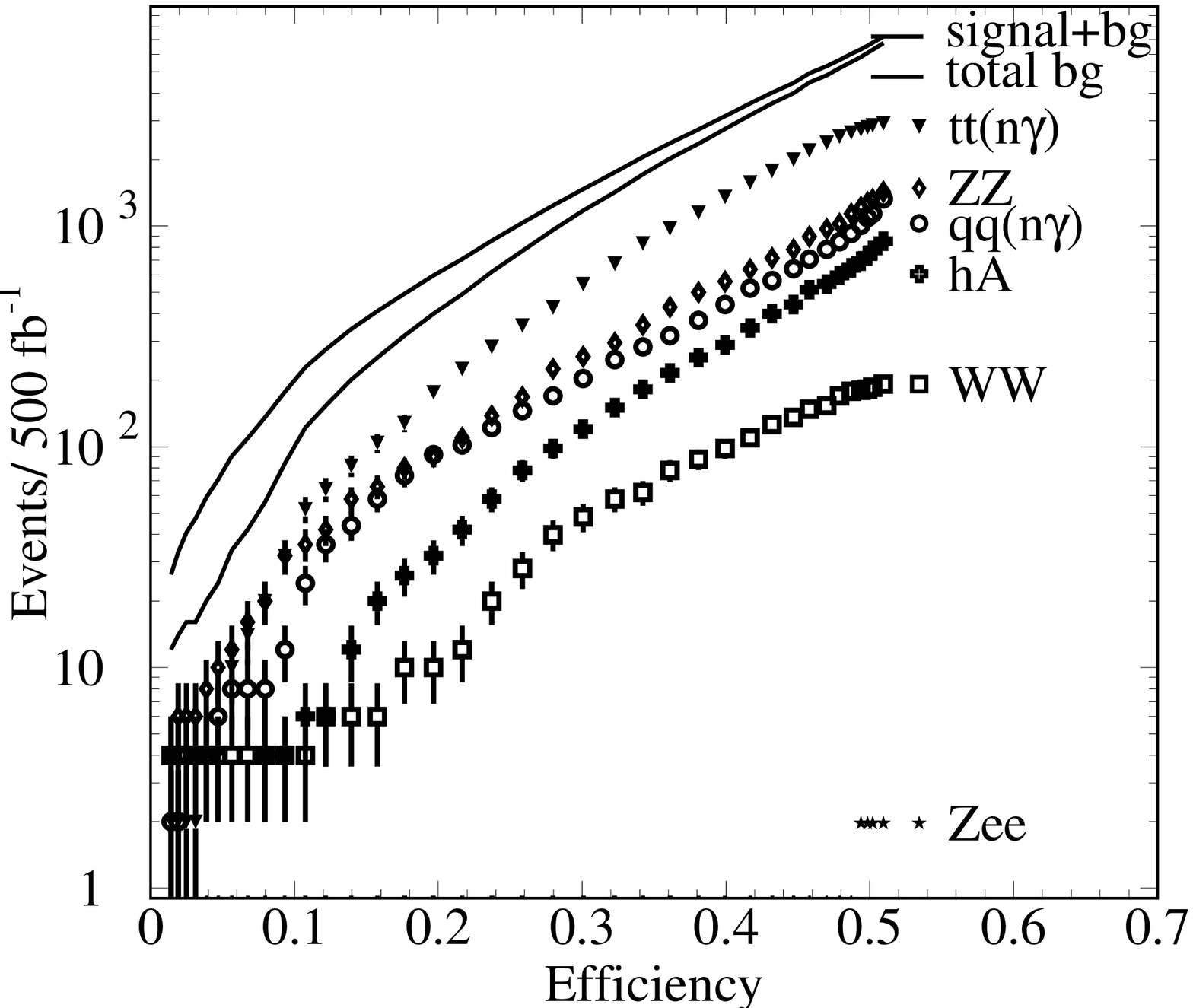,width=\textwidth}}
\end{center}
\end{minipage}
\hfill
\begin{minipage}{0.48\textwidth}
\begin{center}
\mbox{\epsfig{file=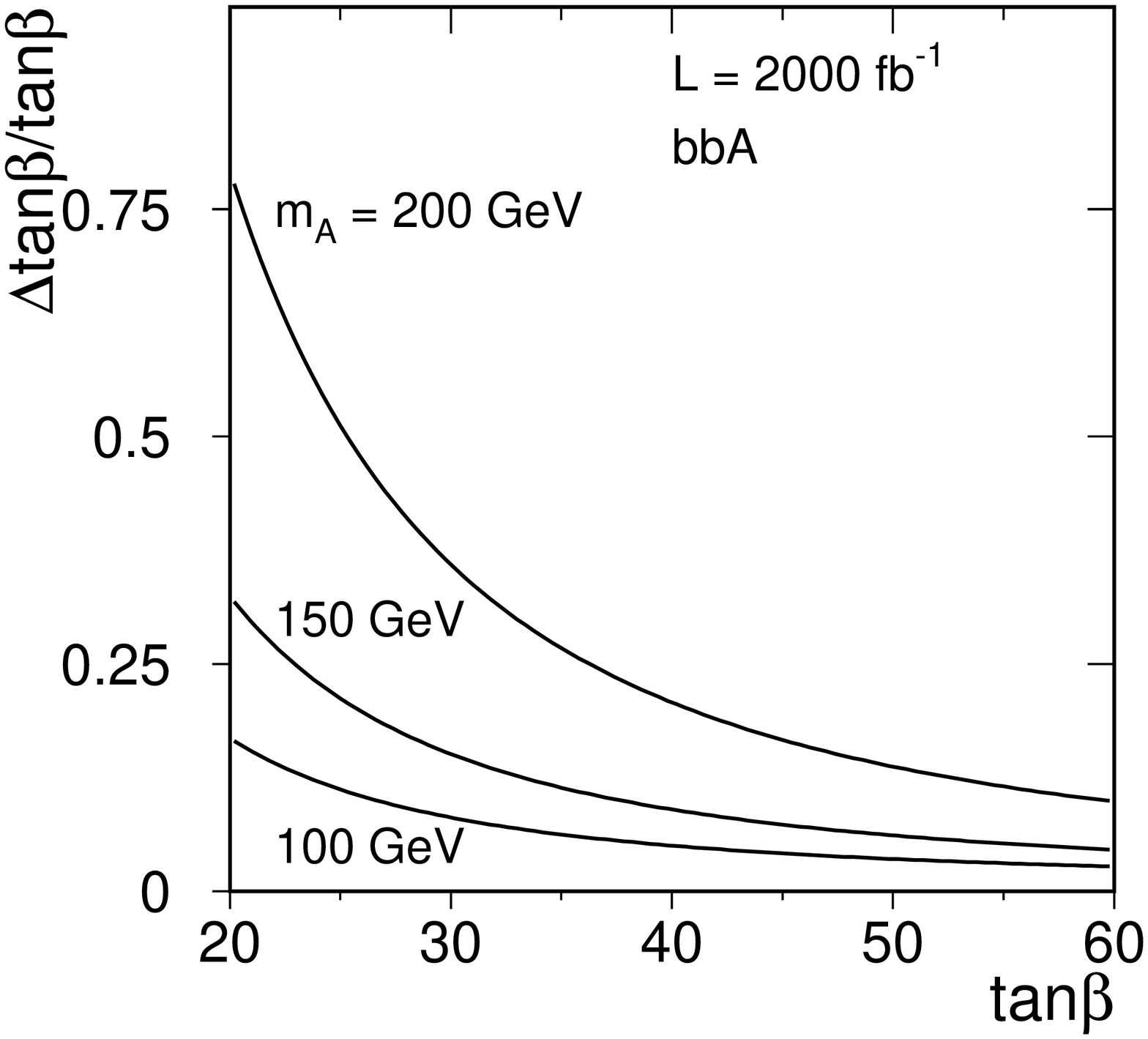,width=\textwidth}}
\end{center}
\end{minipage}
\vspace*{-1cm}
\caption{\label{fig:ida2}
Left: 
Final background rate versus $\bb\ha$ signal efficiency for 
$\mha=100\gev$, $\rts=500\gev$ and $\call=500\fbi$. 
Right:
Corresponding $\tanb$ statistical error for $\call=2000\fbi$
and $\mha=100,150,200\gev$.
For both plots, we take a fixed value of $m_b=4.62\gev$. 
}
\vspace*{-0.2cm}
\end{figure}

The experimental challenge of this study is the low expected production rate 
and the large irreducible background for a four-jet final state, as discussed 
in a previous simulation~\cite{epj}.
The expected background rate for a given $\rm b\anti b\ha\to\bb\bb$ signal
efficiency is shown in Fig.~\ref{fig:ida2}. Taking a working point of
10\% efficiency, we estimate the statistical error in determining $\tanb$ by
$\Delta\tan^2\beta / \tan^2\beta = \Delta S / S
=\sqrt{ S + B} / S =\sqrt{200}/100\approx 0.14,$ 
resulting in an error on $\tanb= 50$ of $7\%$.
In the MSSM, the $\rm \bb\hl$ signal would essentially double the
number of signal events and have exactly the same $\tanb$ dependence,
yielding $\Delta\tan^2\beta/\tan^2\beta\approx \sqrt{300}/200\approx 0.085$
for $\tanb=50$ and the $\tanb$ error would be about $4\%$.
Systematic errors arising from interference with the $\rm hA\ra \bb\bb$
reaction can be controlled~\cite{long}.

\section*{\boldmath $\rm \hh\ha$ production: branching ratios and decay widths}
\vspace*{-6mm}

\begin{figure}[h!]
\vspace*{-5mm}
\begin{center}
\mbox{\epsfig{file=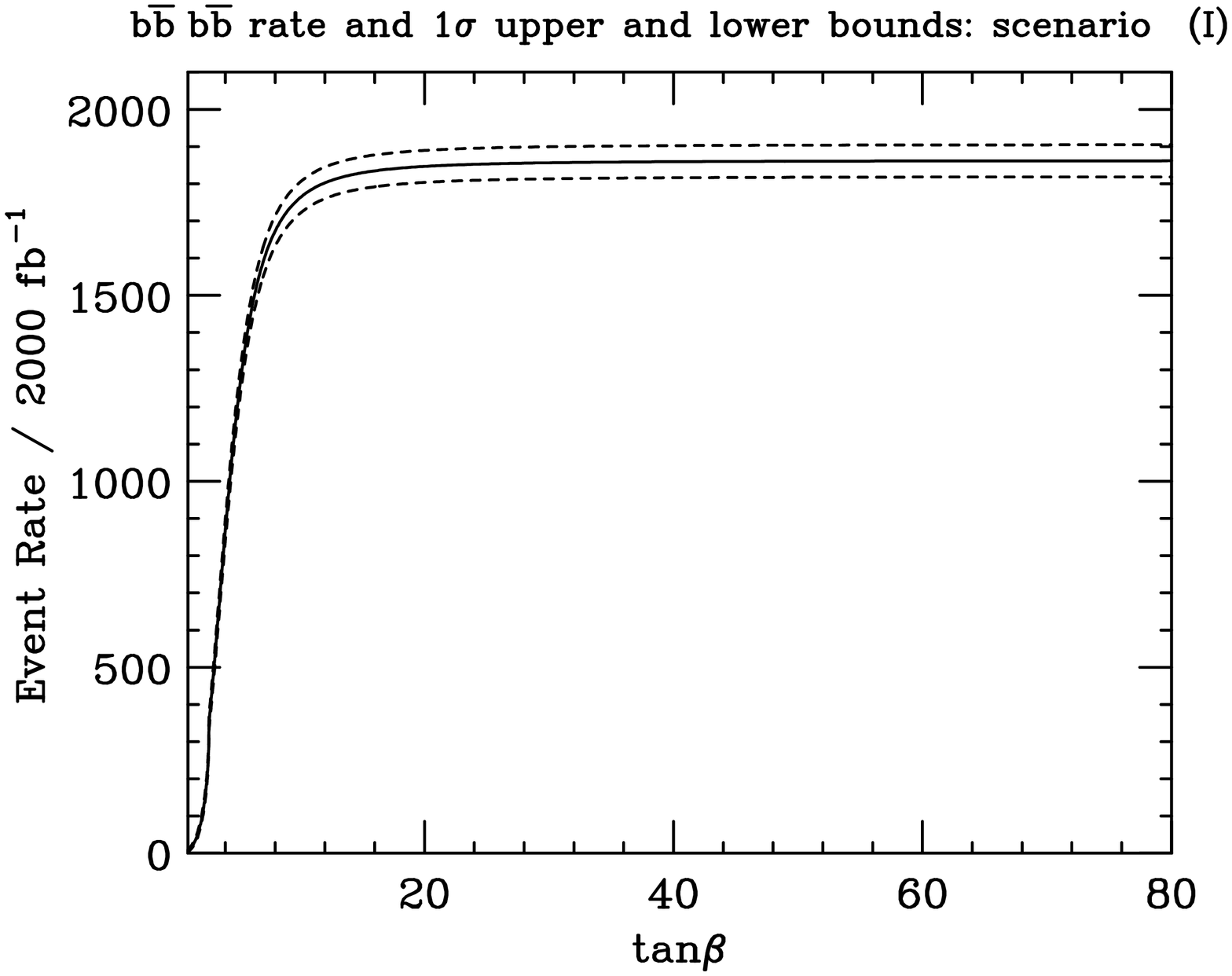,angle=90,width=0.48\textwidth}}
\mbox{\epsfig{file=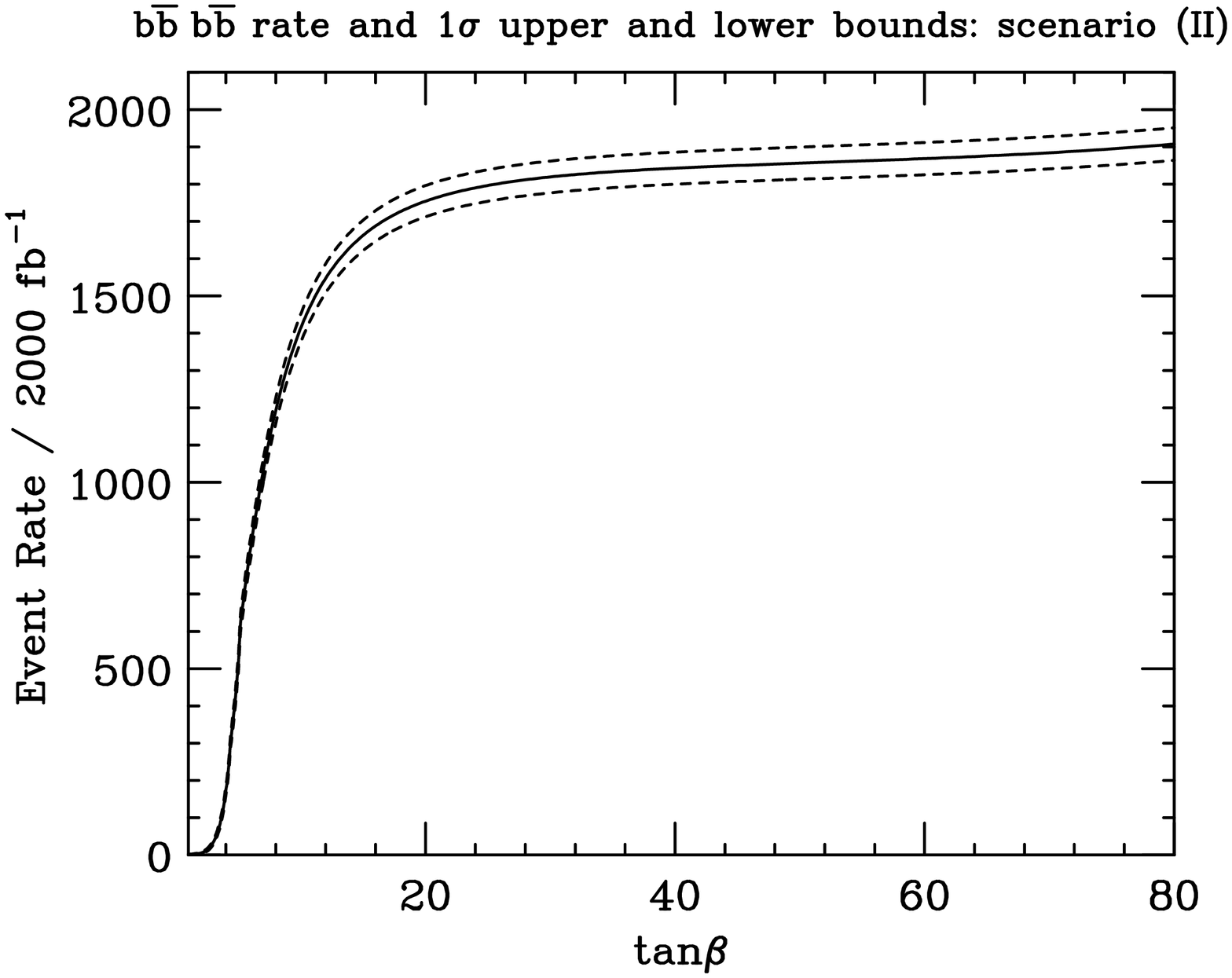,angle=90,width=0.48\textwidth}}
\end{center}
\vspace*{-0.8cm}
\caption{\label{hhhawidrate}
Expected $\rm \epem\to \hh\ha\to \bbbb$ event rates for 10\% efficiency 
and $\pm1\sigma$ bounds in scenarios (I) and (II)
in the MSSM for $\mha=200\gev$, $\rts=500\gev$ and $\call=2000\fbi$.}
\end{figure}

\begin{figure}[h!]
\vspace*{-0.3cm}
\begin{center}
\mbox{\epsfig{file=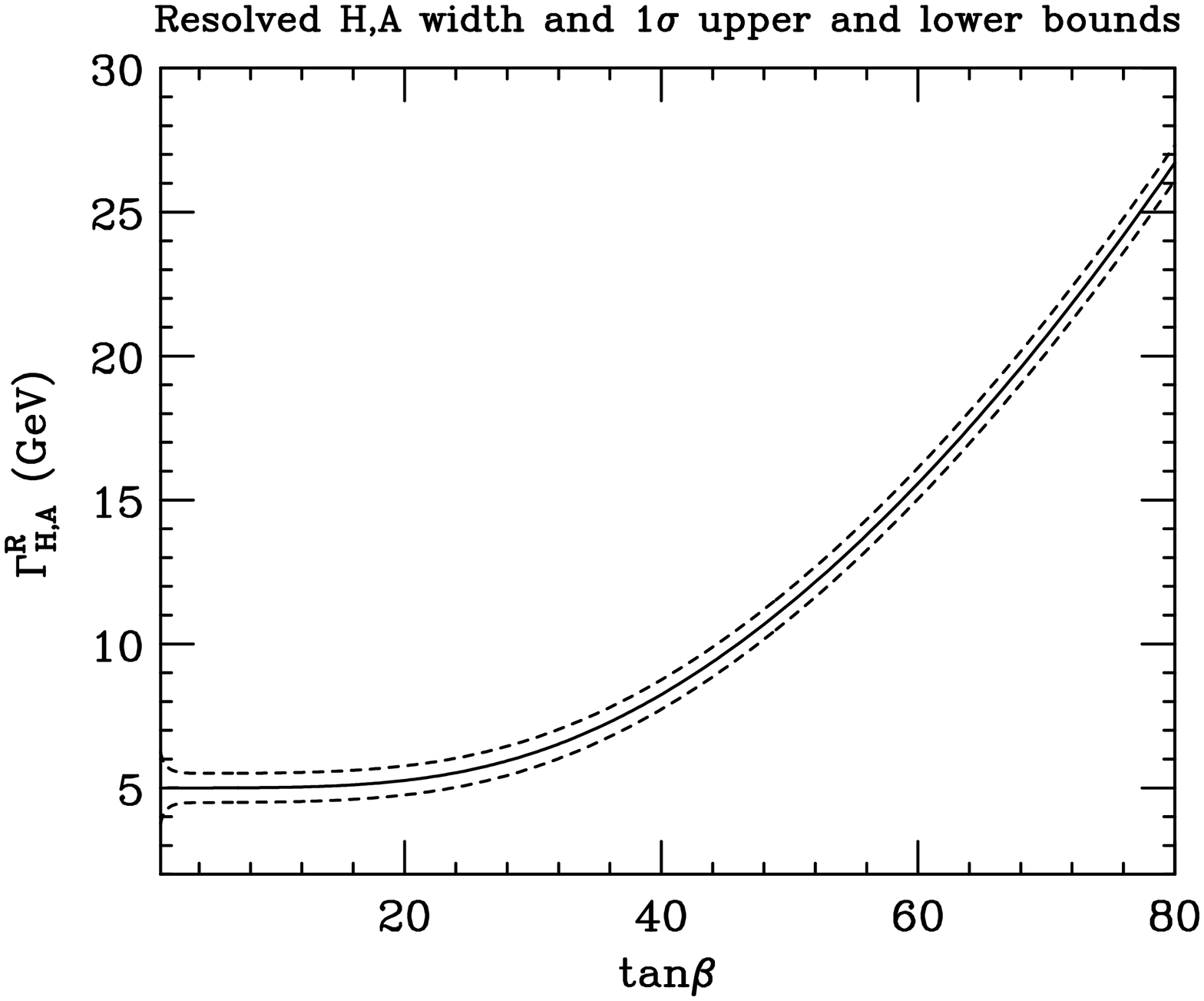,angle=90,width=0.48\textwidth}}
\mbox{\epsfig{file=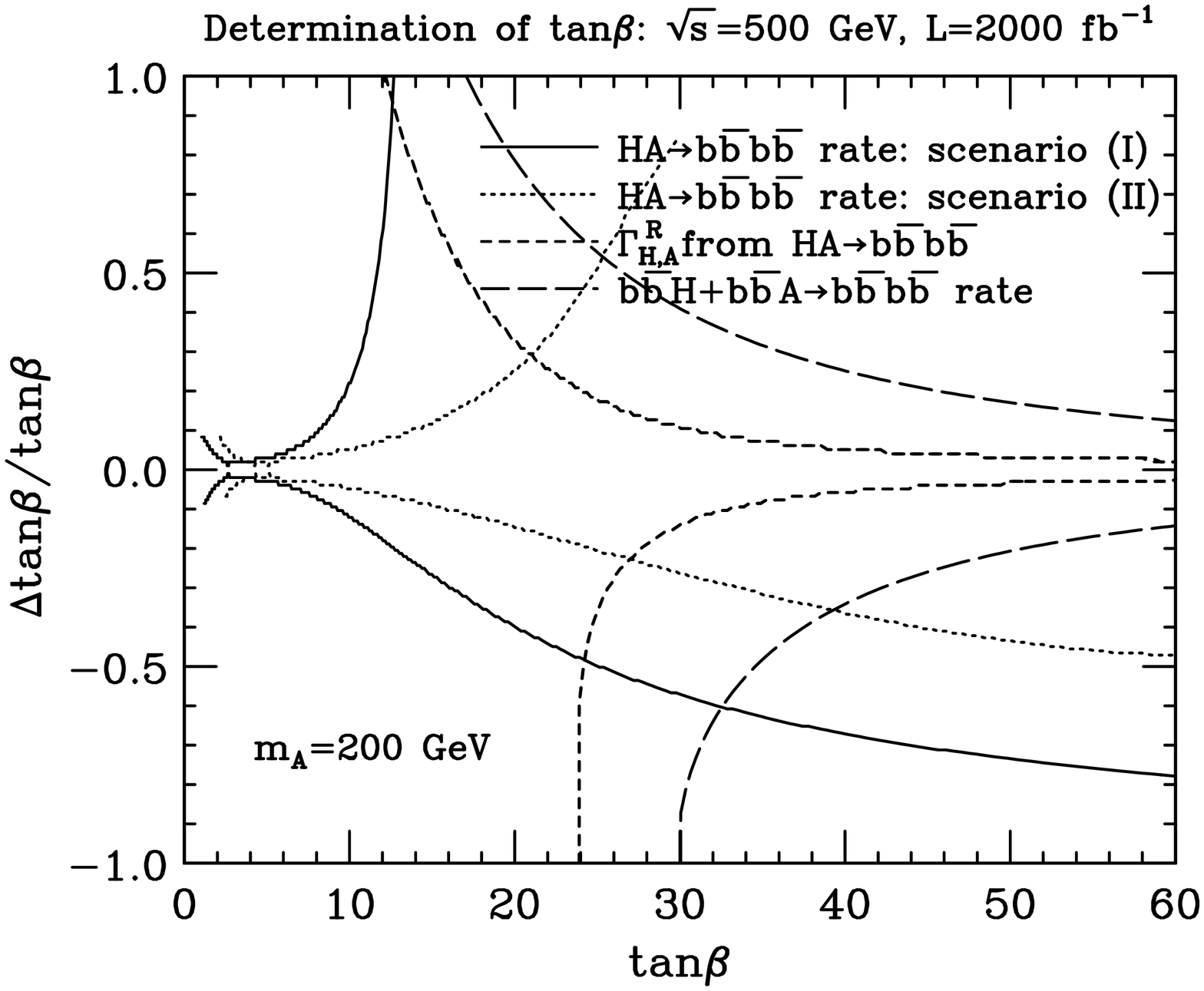,angle=90,width=0.48\textwidth}}
\end{center}
\vspace*{-0.8cm}
\caption{\label{hhhaonly}
Left: Expected resolved width $\Gamma_{\hh,\ha}^{\rm R}$, Eq.~(\ref{gamr}),
for scenario (I) and $1\sigma$ upper and lower bounds with
10\% selection efficiency.
The statistical bounds include an additional efficiency factor of 0.75 
for keeping only events in the central mass peak and assume a 
detector resolution of $\gamres=5\gev$ with a 10\% uncertainty.
Right:
Expected precision on $\tanb$ ($1\sigma$ bounds) for $\mha = 200$~GeV, 
$\rts=500\gev$ and $\call=2000\fbi$ based on
the $\epem\to\bb\ha+\bb\hh\to\bbbb$ rate,
the $\epem\to\hh\ha\to\bbbb$ rate 
and $\Gamma^{\rm R}_{\hh,\ha}$.
}
\vspace*{-0.4cm}
\end{figure}

The branching ratios for $\rm \hh$, $\rm \ha$ decay to various allowed modes
vary rapidly with $\tanb$ in the MSSM when $\tanb$ is below 20.
Consequently, if these branching ratios can be measured accurately,
$\tanb$ can be determined with good precision in this range.
As the H and A decay rates depend on the MSSM parameters,
two cases are considered.
In scenario (I), SUSY decays of the $\rm \hh$ and $\rm \ha$ are kinematically
forbidden. Scenario (II) is taken from~\cite{tao} in which
SUSY decays (mainly to $\wt \chi_1^0\wt\chi_1^0$) are allowed.
We assume event selection criteria with an event selection efficiency 
of $10\%$ and negligible background, based on the expected b-tagging
performance and kinematic event selection.
The expected $\rm \hh\ha\to\bb\bb$ event rates and $1\sigma$ statistical 
bounds
are shown in Fig.~\ref{hhhawidrate} 
as a function of $\tanb$ for $\rts=500\gev$ and $\call=2000\fbi$.
The resulting bounds for $\tanb$
are plotted in Fig.~\ref{hhhaonly} (right)\,for\,MSSM\,scenarios\,(I)\,and\,(II).

A previous HA simulation~\cite{troncon} indicates that about 25\% of 
the time wrong jet-pairings are made, which are attributed to the wings 
of the mass distribution. 
The $m_{\rm \bb}$ values from H and A decays are binned in a 
single distribution, since the H and A mass splitting is
typically substantially smaller than the detector resolution 
of $\gamres=5\gev$ for the large $\tanb$ values considered.
Our effective observable is the resolved average width defined by
\vspace*{-0.1cm}
\beq
\Gamma^{\rm R}_{\rm \hh,\ha}=\half\left[\sqrt{[\gamhhtot]^2+[\gamres]^2}+
\sqrt{[\gamhatot]^2+[\gamres]^2}\right].
\label{gamr}
\eeq
Its dependence on $\tan\beta$ is shown in Fig.~\ref{hhhaonly}
for $\mhh\approx\mha=200\gev$ in MSSM scenario (I)
and it is very similar for scenario (II).

In order to extract the implied $\tanb$ bounds,
we must account for the fact that
the detector resolution will not be precisely determined.
There will be a systematic uncertainty
which we have estimated at 10\% of $\Gamma_{\rm res}$, i.e. 0.5 GeV.
This systematic uncertainty considerably weakens our ability to
determine $\tanb$ at the lower values of $\tanb$ for which
$\rm \gamhhtot$ and $\gamhatot$ are smaller than $\Gamma_{\rm res}$. This
systematic uncertainty
should be carefully studied as part of future experimental analyses.
Figure~\ref{hhhaonly} shows also the expected $\pm 1\sigma$ experimental
errors based on the measurement of $\Gamma^{\rm R}_{\rm \hh,\ha}$.
An excellent determination of $\tanb$ will be possible at high $\tanb$.
The $\rm b\bar bH/A$ and H/A width methods are nicely complementary in
their $\tanb$ coverage to the $\tanb$ determination based on the
$\rm \hh\ha\to\bb\bb$ rate method at lower $\tanb$.

\vspace*{-0.6cm}
\section*{\boldmath $\rm H^+H^-$ production: branching ratios and decay widths}
\vspace*{-0.6cm}

The reaction $\rm \epem\to H^+H^-\to \tbtb$ can be 
observed at a LC~\cite{as:hphm500} and recent high-luminosity 
simulations~\cite{lc800:kiiskinen} show that precision measurements
can be performed. As soon as the charged
Higgs boson decay into $\rm tb$ is allowed this decay mode is dominant.
Nonetheless, $\rm \br(\hpm\to tb)$ varies significantly with $\tan\beta$, 
especially for small values of $\tanb$ where the $\rm tb$ mode competes with
the $\tau\nu$ mode.
The $\rm H^+\to t\anti b$ branching ratio and width are sensitive to $\tanb$
in the form
${\Gamma(\hpm\to tb)} \propto \mt^2 \cot^2\beta + \mb^2\tan^2\beta\,.$
As in the previous section, we use HDECAY~\cite{hdecayref}  
(which incorporates the running of the b-quark mass) 
to evaluate the charged Higgs boson branching ratios and decay widths.
The $\rm tb$ partial width and the corresponding branching ratio have a 
minimum in the vicinity of $\tanb\approx 6-8$. 
In contrast to the variation of the branching ratio,
the cross section for $\epem\to\hp\hm$ production
is largely independent of $\tan\beta$.

\begin{figure}[h!]
\begin{center}
\mbox{\epsfig{file=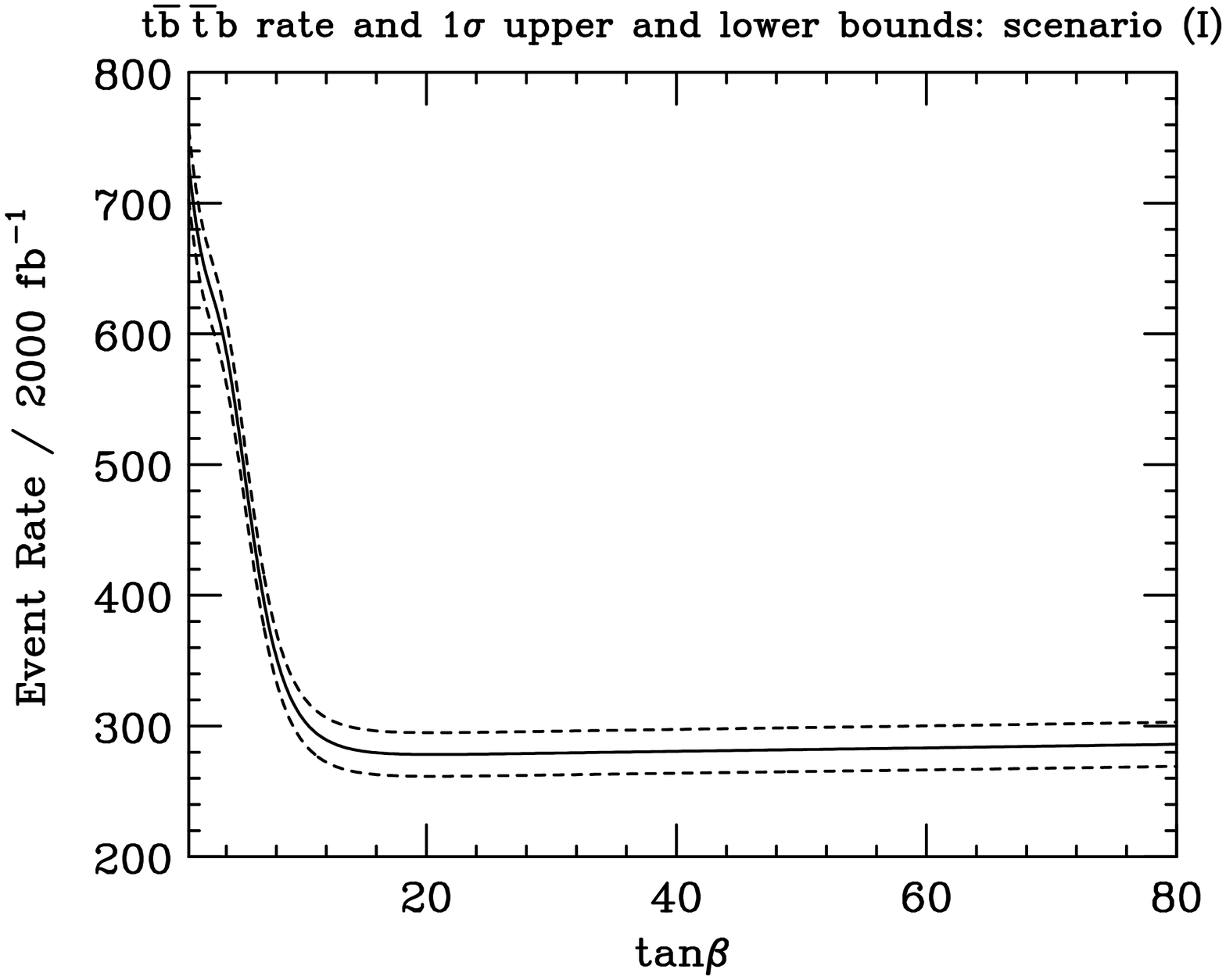,angle=90,width=0.48\textwidth}}
\mbox{\epsfig{file=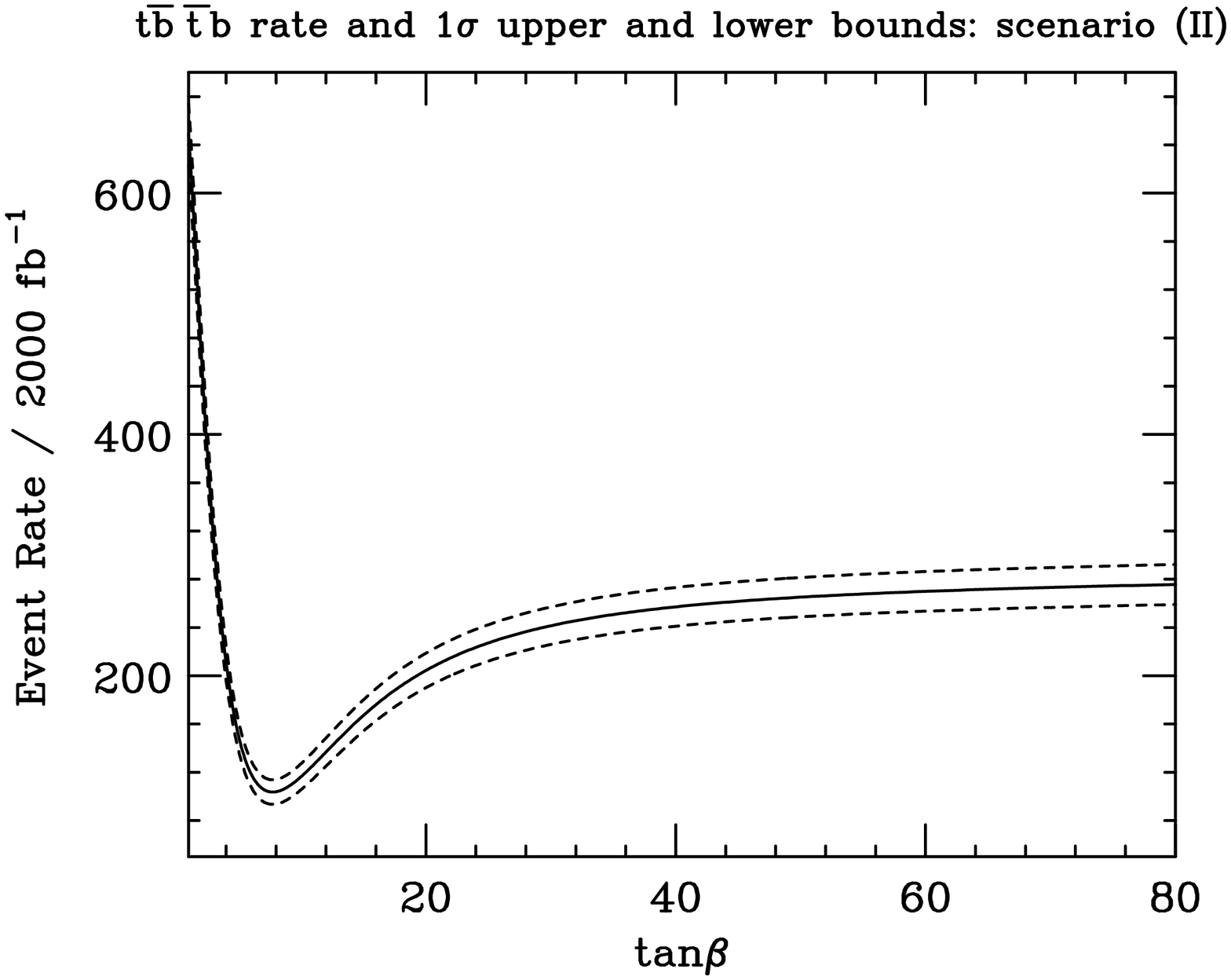,angle=90,width=0.48\textwidth}}
\end{center}
\vspace*{-0.8cm}
\caption{\label{hpmwidrate}
Expected $\rm \epem\to \hp\hm\to \tbtb$ event rates for 2.2\% 
efficiency and $\pm1\sigma$ bounds in scenarios (I) and (II)
in the MSSM for $\mha=200\gev$, $\rts=500\gev$ and $\call=2000\fbi$.
}
\end{figure}

\begin{figure}[h!]
\vspace*{-3mm}
\begin{center}
\mbox{\epsfig{file=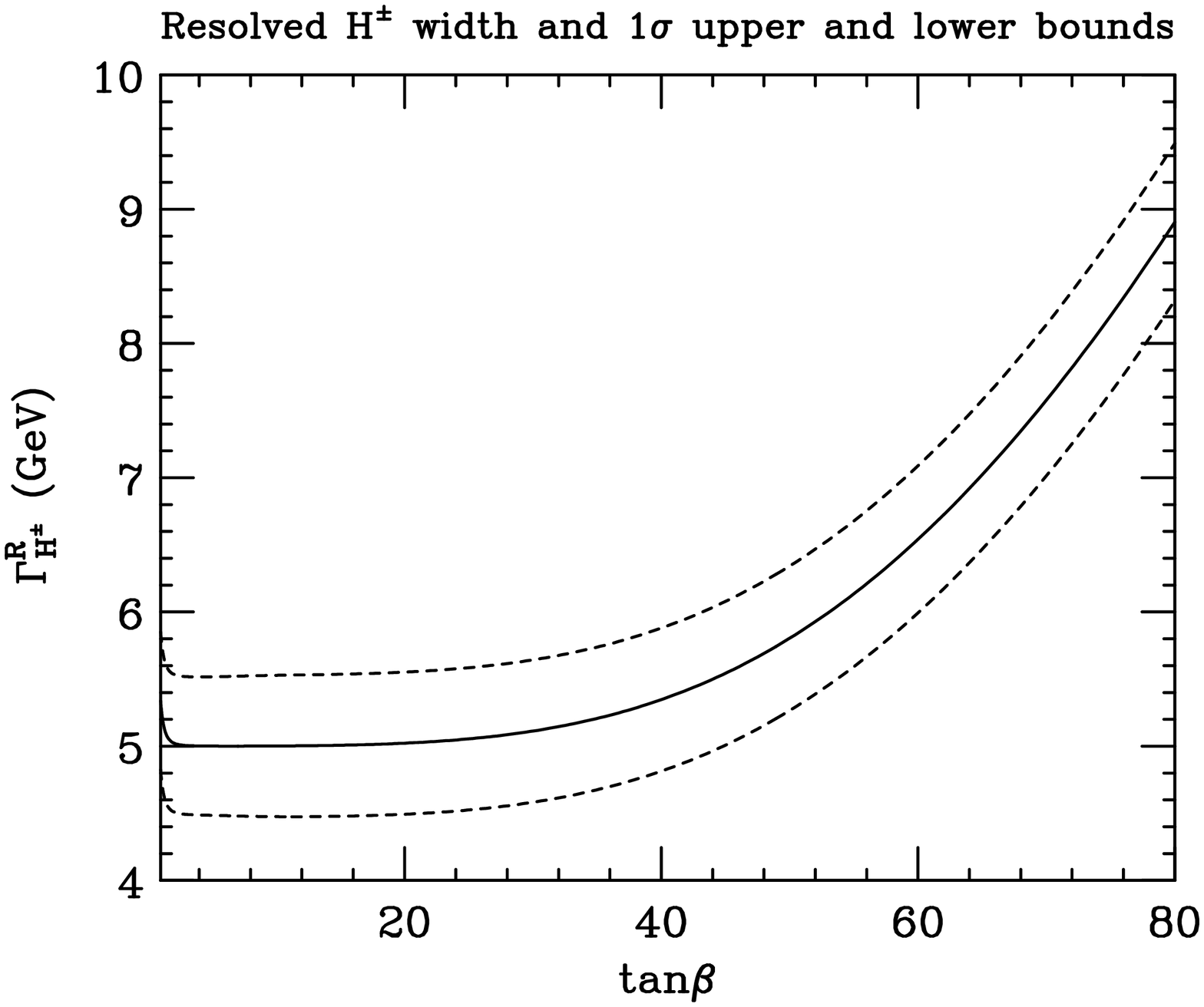,angle=90,width=0.48\textwidth}}
\mbox{\epsfig{file=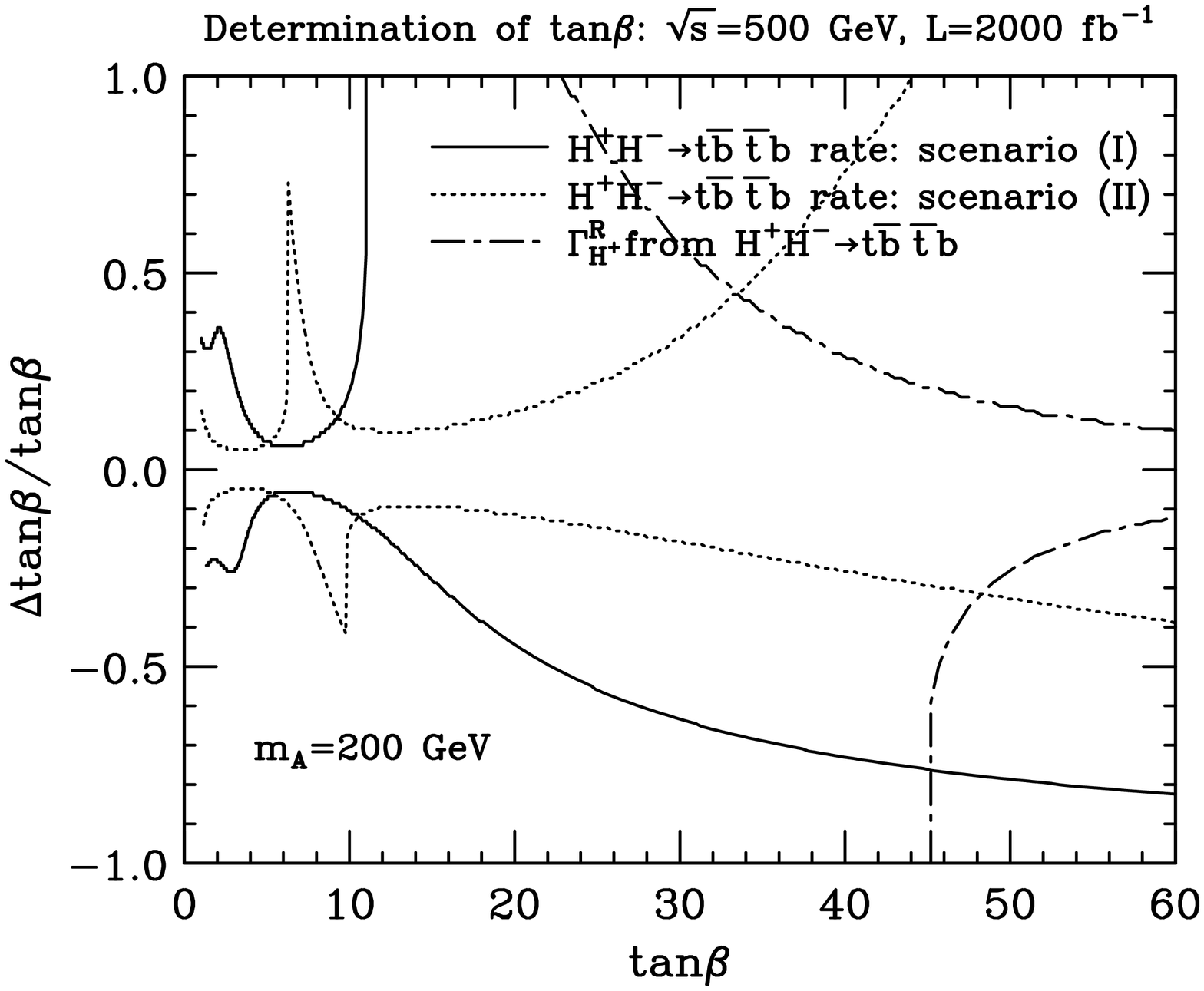,angle=90,width=0.48\textwidth}}
\end{center}
\vspace*{-0.8cm}
\caption{\label{hpmonly} 
Left: Expected resolved width $\Gamma_{\hpm}^{\rm R}$, Eq.~(\ref{gamrhpm}),
for scenario (I) and $1\sigma$ upper and lower bounds with
2.2\% selection efficiency.
The statistical bounds include an additional efficiency factor of 0.75 
for keeping only events in the central mass peak and assume 
$\gamres=5\gev$ with a 10\% uncertainty.
Right:
Expected precision on $\tanb$ ($1\sigma$ bounds) 
for $\mhpm\approx\mha = 200$~GeV, 
$\rts=500\gev$ and $\call=2000\fbi$ based on
the $\epem\to\hp\hm\to\tbtb$ rate and
$\Gamma^{\rm R}_{\hpm}$.
}
\end{figure}

Our procedures for estimating errors for the $\tbtb$
rate and for the total width 
are similar to those given earlier for $\rm \hh\ha$ production
rate and width in the $\rm \bbbb$ channel.
For $\mhpm=300\gev$ at $\rts=800\gev$, a $\rm H^+H^-$ 
study~\cite{lc800:kiiskinen}
finds that the $\tbtb$ final state can be isolated with an efficiency 
of $2.2\%$. For $\mhpm=200\gev$ and $\rts=500\gev$,  
we have adopted the same $2.2\%$ efficiency and negligible background.
Figure~\ref{hpmwidrate} shows the resulting $\tbtb$ rates
and $1\sigma$ bounds for MSSM scenarios (I) and (II). 
The corresponding bounds on $\tan\beta$ are shown 
in Fig.~\ref{hpmonly} (right).

For the total width determination, we assume that we
keep only 75\% of the events after cuts (\ie\
a fraction $0.75\times 0.022$ of the raw event number), corresponding
to throwing away wings of the mass peaks, and each $\tbtb$ event
is counted twice since we can look at both the $\rm \hp$ and the $\rm \hm$
decay. We define a resolved width which incorporates the detector 
resolution 
$\rm \gamres=5\gev$:
\beq
\Gamma^{\rm R}_{\hpm}=\sqrt{[\gamhpmtot]^2+[\gamres]^2}.
\label{gamrhpm}
\eeq
Estimated errors are based on the width measurement for 10\% systematic
error in $\gamres = 0.5\gev$.
The resolved width $\Gamma^{\rm R}_{\hpm}$ for scenario (I)
is given in Fig.~\ref{hpmonly} and it is very similar for scenario (II).
It also shows resulting $\tanb$ bounds.
In comparison to the neutral Higgs boson methods (Fig.~\ref{hhhaonly}),
we observe that for MSSM scenario (I) the $\tbtb$ rate measurement gives a 
$\tanb$ determination that is quite competitive with that from $\hh\ha$
production in the $\bbbb$ final state.  For MSSM scenario (II),
the $\tbtb$ rate gives an even better $\tanb$ determination
than does the $\bbbb$ rate. On the other hand,
the width measurement from the $\tbtb$ final state of $\hp\hm$
production is much poorer
than that from the $\bbbb$ final state of $\hh\ha$ production.

\begin{figure}[h!]
\begin{minipage}{0.5\textwidth}
\begin{center}
\mbox{\epsfig{file=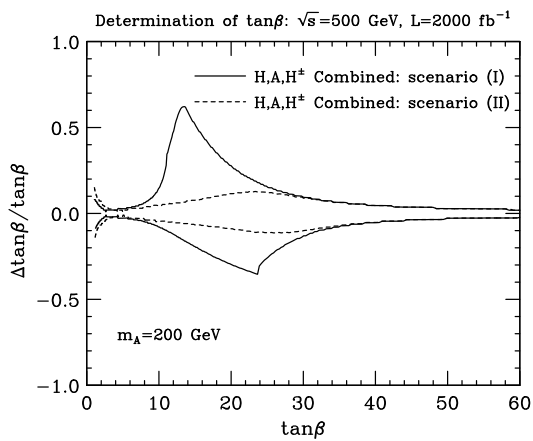,angle=90,width=1.0\textwidth}}
\end{center}
\end{minipage}\hfill
\begin{minipage}{0.4\textwidth}
\caption{\label{totalonly} 
Expected combined precision on $\tanb$ ($1\sigma$ bounds)
for $\mhpm\approx \mha = 200$~GeV,
$\rts=500\gev$ and $\call=2000\fbi$ based on
combining (in quadrature) the results shown in 
Figs.~\ref{hhhaonly} and \ref{hpmonly}
for the $\rm b\bar bH/A$ rate, 
the $\rm HA\to\bb\bb$ rate, 
$\Gamma_{\hh,\ha}^{\rm R}$, 
the $\tbtb$ rate and 
$\Gamma^{\rm R}_{\hpm}$.
Higher order calculations in the MSSM could
influence the combination of the different $\tanb$ methods.
}
\end{minipage}
\vspace*{-1.0cm}
\end{figure}

By combining the $\tanb$ errors from all processes in quadrature
we obtain the expected net errors on $\tanb$ shown in Fig.~\ref{totalonly}
for MSSM scenarios (I) and (II).
The Higgs sector will provide an excellent determination
of $\tanb$ at small and large $\tanb$ values.
However, larger bounds are expected for moderate $\tanb$ in scenario (II)
where SUSY decays of the $\ha,\hh,\hpm$ are not significant.
Further information on $\tanb$ could be obtained from the reaction
$\rm \epem\to t\bar t\to tbH^\pm\to tb\tau\nu$, further Higgs decay branching
ratios (e.g. $\rm H\to WW,ZZ,hh;~A\to Zh;~H,A\to SUSY$ particles), 
the H/A decay width from $\rm b\bar b H/A$ production, and the polarization
of scalar taus.

\vspace*{-0.5cm}
\section*{Conclusions}
\vspace*{-0.6cm}

A high-luminosity linear $\rm e^+e^-$ collider will provide 
a precise measurement of the value of $\tanb$
throughout most of the large possible $\tan\beta$ range $1<\tanb<60$.
In particular, we have demonstrated the complementarity
of employing: 
a) the $\bb\ha+\bb\hh \to \bbbb$ rate; 
b) the $\hh\ha\to \bbbb$ rate; 
c) the average $\hh,\ha$ total width from $\hh\ha$ production;
d) the $\hp\hm\to \tbtb$ rate; and 
e) the $\hpm$ total width from $\hp\hm\to\tbtb$ production.
Experimental challenges will be the required high total luminosity,
an excellent b-tagging performance, and precision detector resolution
and selection efficiency determinations.


\end{document}